\documentclass[epj,referee]{svjour}
\usepackage{graphics}
\usepackage{graphicx}
\usepackage{amsmath}
\usepackage{amssymb}
\usepackage{epsfig}
\usepackage{dsfont}
\usepackage{xcolor}
\begin{document}
\title{An observable prerequisite for the existence of persistent currents}
\author{Jacob~Szeftel\inst{1} \and Nicolas~Sandeau\inst{2}\and Michel Abou Ghantous\inst{3}}
\institute{ENS Cachan, LPQM, 61 avenue du Pr\'esident Wilson, 94230 Cachan \and Aix Marseille Univ, CNRS, Centrale Marseille, Institut Fresnel, F-13013 Marseille, France \and American University of Technology, AUT Halat, Highway, Lebanon}
\abstract{A classical model is presented for persistent currents in superconductors. Their existence is argued to be warranted because their decay would violate the second law of thermodynamics. This conclusion is achieved by analyzing comparatively Ohm's law and the Joule effect in normal metals and superconducting materials. Whereas Ohm's law applies in identical terms in both cases, the Joule effect is shown to cause the temperature of a superconducting sample to \textit{decrease}. An experiment is proposed to check the validity of this work in superconductors of both types I and II.}
\PACS{{74.20.Mn}{Nonconventional mechanisms}\and{74.25.Bt}{Thermodynamic properties}\and{74.25.Fy}{Transport properties}}
\maketitle
		\section{Introduction}
	The prominent signature of superconductivity, i.e. the property to sustain persistent currents\cite{ash,par,tin} in vanishing electric field, has remained unexplained, since its discovery\cite{kam}, as stressed by Ashcroft and Mermin\cite{ash} (see\cite{ash} p.750, $1^{st}$ paragraph, line $1$) : \textit{the property for which the superconductors are named is the most difficult to extract from the microscopic} (i.e. BCS\cite{bcs}) \textit{theory}. This is all the more disturbing, since the mainstream narrative on superconductivity relies heavily on the phenomenological equations, proposed by London\cite{lon} and Ginzburg and Landau\cite{gin}, for which the existence of persistent currents is merely \textit{assumed}.\par 
	 In order to understand why this long-standing riddle has withstood every attempt\cite{way} at elucidating it so far, it is helpful to recall the basic tenets of electric conductivity in normal conductors\cite{ash}. The applied electric field $E$ accelerates the electrons in the conduction band, which gives rise to a current $j$. Eventually, the driving force $\propto E$ is counterbalanced by a friction one $\propto j$, exerted by the lattice, as conveyed by Ohm's law :
\begin{equation}
\label{ohm}
 j=\sigma E\quad,\quad \sigma=\frac{c_0e^2\tau}{m}\quad,
\end{equation}
where $\sigma,c_0,e,m,\tau$ stand for the conductivity, the electron concentration, the electron charge, its effective mass, and the decay time of $j$ due to friction, respectively. Simultaneously, the work performed by the electric force is entirely transformed into heat, to be released in the lattice, through the Joule effect. As a consequence of Eq.(\ref{ohm}), the observation\cite{kam} of $j\neq 0$ despite $E=0$ seemed indeed to suggest $\tau\rightarrow\infty\Rightarrow\sigma\rightarrow\infty$. However, it is well-known nowadays that both $\tau,\sigma$ are finite, provided the measurement is carried out with an \textit{ac} current, as emphasized by Schrieffer\cite{sch} (see\cite{sch} p.4, $2^{nd}$ paragraph, lines $9,10$): \textit{at finite temperature, there is a finite ac resistivity for all frequencies $>0$}. For instance, the conductivity, measured in $YBa_2Cu_30_7$ below the critical temperature $T_c$, has been reported\cite{ges,sar} to be such that $\sigma\approx 10^5\sigma_n$, where $\sigma_n$ stands for the normal conductivity, measured just above $T_c$. Additional evidence is provided by commercial microwave cavity resonators, made up of superconducting materials, displaying a very high, albeit \textit{finite} conductivity (see\cite{tin} lowest line in p.38). Finally, the observable consequences of finite $\sigma$, regarding the skin\cite{jac,bor} and Meissner\cite{lon,mei} effects, have been discussed recently\cite{sz1,sz2} and the finite resistivity of superconductors has been ascribed \textit{solely} to \textit{superconducting} electrons\cite{sz3} on the basis of susceptibility data.\par 
	Therefore the issue of persistent currents will be tackled here from quite different a starting point. Likewise we shall  show how the very properties of the BCS state\cite{bcs} cause the Joule dissipation to be thwarted in a superconductor, undergoing no electric field. This goal will be achieved by making a comparative study of Ohm's law and the Joule effect in normal and superconducting metals, based on Newton's law and the two laws of thermodynamics.\par 
	 The outline is as follows : the conditions for a superconductor to be in thermal equilibrium are discussed in sections 2, while stressing the different properties of the BCS state\cite{bcs} versus those of the Fermi gas\cite{ash}; Ohm's law and the Joule effect are studied in sections 3 and 4, respectively; a necessary condition for  the existence of persistent currents is worked out in section 5, while an experiment, enabling one to check the validity of this analysis in superconducting materials of both kinds, is described in section 6. Our observable predictions will turn out to concur very well with a remark by De Gennes\cite{gen}. The results of this work are summarized in the conclusion. 
		\section{The two-fluid model}
	The conduction properties of a superconducting material will be analyzed within the two-fluid model\cite{par,tin,sch,gen}. In this framework, the conduction electrons make up a homogeneous mixture, in thermal equilibrium, of normal and superconducting electrons, in concentration $c_n,c_s$, respectively.\par
	  All of the electronic properties of the normal state are governed by the Fermi-Dirac statistics, and thence accounted for within the Fermi gas\cite{ash} model. In particular, its Helmholz free energy per unit volume $F_n$ depends on two parameters, the temperature $T$ and the Fermi energy $E_F$, defined\cite{ash,lan} as the chemical potential of independent electrons, i.e. $E_F=\frac{\partial F_n}{\partial c_n}$.\par
	 By contrast, the BCS wave-function\cite{bcs} describes the motion of superconducting electrons, as a \textit{many-body bound} state, which entails that the BCS energy per unit volume $\mathcal{E}_s$ depends \textit{only} on the concentration of superconducting electrons $c_s$. Because $\mathcal{E}_s$ is $T$-independent, the BCS state\cite{bcs}, unlike the Fermi gas, is inferred to carry no entropy\cite{ash,par,tin}, so that its free energy is equal to $\mathcal{E}_s$ (this property is confirmed experimentally by the weak thermal conductivity\cite{ash,par,tin}, measured in superconductors, in marked contrast with the high one, typical of normal metals). Thus the chemical potential $\mu$ of the BCS state reads $\mu=\frac{\partial \mathcal{E}_s}{\partial c_s}$.\par
	The equilibrium, achieved in the two-fluid model, stems from Gibbs and Duhem's law\cite{lan}, which requires the free energy of the whole electron system $F_e=F_n(T,c_n)+\mathcal{E}_s(c_s)$ to be minimum with respect to $c_n,c_s$, under the constraints $c_n+c_s=c_0$ ($c_0$ refers to the total concentration) and $T$ kept constant, and thence leads to
\begin{equation}
\label{gidu}
E_F(T,c_n)=\mu(c_s)\quad.
\end{equation}
 The peculiar properties of the Joule effect, taking place in a BCS state, will appear below to be solely determined by the \textit{sign} of $\frac{\partial \mu}{\partial c_s}=\frac{\partial^2 \mathcal{E}_s}{\partial c_s^2}$.\par
	 An early, phenomenological attempt\cite{cas}, aimed at explaining the specific heat data, measured in superconducting materials, made use of Eq.(\ref{gidu}) too. However our approach differs from that one, inasmuch as it refrains from assuming specific, but arbitrary expressions for $F_n(T,c_n),\mathcal{E}_s(c_s)$, so that our conclusions do not suffer from any loss of generality. 
\begin{figure}
\includegraphics*[height=5 cm,width=5 cm]{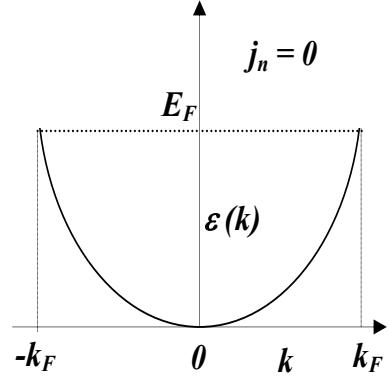}
\caption{schematic projected one electron dispersion $\epsilon(k)$ of occupied states $(\epsilon(k)\in\left[0,E_F\right])$ for $j_n=0$ as a solid line}\label{o1}
\end{figure}
\begin{figure}
\includegraphics*[height=5 cm,width=5 cm]{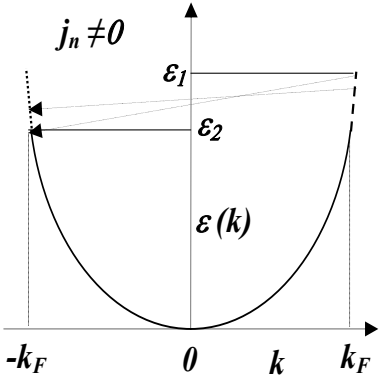}
\caption{schematic projected one electron dispersion $\epsilon(k)$ for $j_n\neq 0$ in the laboratory frame; the solid line represents most of electrons, which contribute nothing to $j_n$, whereas the dashed line corresponds to the few electrons responsible for $j_n\neq 0$; actually $j_n$ is proportional to the length of the dashed line but the tiny difference $\epsilon_1-\epsilon_2$ has been hugely magnified for the reader's convenience, that is $\epsilon_1\approx \epsilon_2\approx E_F$; the dotted line represents empty one electron states; the arrows illustrate electron transitions, from occupied (dashed line) back to empty (dotted line) states, driven by the friction force}\label{o2}
\end{figure}
	\section{Ohm's law}
 Owing to Fermi-Dirac statistics and $T<<T_F=\frac{E_F}{k_B}\approx3\times 10^4K$ ($k_B$ stands for the Boltzmann constant), the electrons in a normal metal make up a degenerate Fermi gas\cite{ash}, for which each one-electron state, with energy ranging from the bottom of the conduction band up to $E_F$, is doubly occupied (due to the two spin directions), whereas those states with energy $>E_F$ remain empty. The corresponding one electron dispersion curve $\epsilon(k)$ has been projected onto the direction of the applied electric field $E$, as pictured in Fig.\ref{o1}. Since the electron velocity\cite{ash} is  equal to $\frac{\partial \epsilon(k)}{\hbar\partial k}$ and thanks to $\epsilon(k)=\epsilon(-k)$, the resulting current $j_n$ vanishes.\par
  The applied field $E$ arouses a finite current $j_n\neq 0$ by accelerating $\delta c_n$ of electrons $(\delta c_n<<c_n)$ from their initial wave-vector $-k_F$ up to their final one $k_F$, with $k_F$ being such that $\epsilon(k_F)=E_F$, due to Pauli's principle. Therefore all electrons, contributing to $j_n$, have about the \textit{same} velocity $v_F=\frac{\partial \epsilon(k_F)}{\hbar\partial k}$, so that the resulting current reads $j_n=2\delta c_n ev_F$ (see the dashed line in Fig.\ref{o2}). Inversely, the friction force, exerted by the lattice on those electrons making up $j_n$, tends to bring $2\frac{\delta c_n}{\tau_n}$ of electrons per unit time from $k_F$ back to $-k_F$, where $\tau_n$, showing up in Eq.(\ref{ohm}) as $\tau$, represents the average time between two successive scattering events\cite{ash} (see the arrows pointing to the dotted line in Fig.\ref{o2}). As the momentum change rate, involved in this process, is equal to
 	$$\frac{\delta p}{\tau_n}=-2\frac{mv_F\delta c_n}{\tau_n}=-\frac{m}{e\tau_n}j_n\quad,$$
  Newton's law reads\cite{sz1,sz2} finally
 \begin{equation}
\label{newt}
\frac{m}{e}\frac{dj_n}{dt}=c_n eE-\frac{m}{e\tau_n}j_n\quad.
\end{equation}
Because the inertial force $\frac{m}{e}\frac{d j_n}{d t}$ has been shown to be negligible\cite{sz1,sz2}, the electric force $c_n eE$ and the friction one $-\frac{m}{e\tau_n}j_n$ cancel each other, so that Eq.(\ref{newt}) boils down to Ohm's law, as expressed in Eq.(\ref{ohm}).\par
 	Ohm's law will be worked out now for a superconductor by proceeding similarly as hereabove. The $j_s=0$ superconducting state ($j_s$ refers to the superconducting current) is assumed to consist in two subsets, each of them comprising the same number of electrons. It ensues, from the very properties of the BCS state\cite{bcs}, flux quantization and Josephson's effect\cite{ash,par,jos}, that the electrons in each subset are organized in pairs, moving in opposite directions with respective velocity $v_s,-v_s$, which ensures $j_s=0$. The driving field $E$ causes $\delta c_s$ of electrons $(\delta c_s<<c_s)$ to be transferred from one subset to the other, which results into a finite current $j_s=2\delta c_s ev_s$. The friction force is responsible for the reverse mechanism, whereby an electron pair is carried from the majority subset of concentration $c_s+\delta c_s$ back to the minority one of concentration $c_s-\delta c_s$. Hence if $\tau_s^{-1}$ is defined as the transfer probability per time unit of one electron pair, the electron transfer rate is equal to 
 		$$\frac{c_s+\delta c_s-(c_s-\delta c_s)}{\tau_s}=2\frac{\delta c_s}{\tau_s}\quad.$$
 	 Then Newton's law reads similarly as Eq.(\ref{newt}), valid for independent electrons
 \begin{equation}
\label{newt2}
\frac{m}{e}\frac{dj_s}{dt}=c_s eE-\frac{m}{e\tau_s}j_s\quad.
\end{equation}
	As for independent electrons, the electric force $c_s eE$ and the friction force $-2\frac{mv_s\delta c_s}{\tau_s}=-\frac{mj_s}{e\tau_s}$ cancel each other, which yields the searched result, identical to Eq.(\ref{ohm})
\begin{equation}
\label{ohm2}
c_s eE=\frac{m}{e\tau_s}j_s\Rightarrow j_s=\sigma_s E\quad,\quad\sigma_s=\frac{c_se^2\tau_s}{m}\quad.
\end{equation}
  Although Ohm's law displays the same expression for normal and superconducting metals as well, it should be noted that $\tau_s>>\tau_n$\cite{ges,sar}.\par
Finally note  that the inter-electron forces, responsible for the binding energy of the BCS state with respect to the corresponding Fermi gas of same electron concentration and also for the two-electron scattering within the Fermi gas, do not show up in Eqs.(\ref{newt},\ref{newt2}). In order to understand this feature, let us consider two electrons labelled $i,j$. They exert the forces $\textbf{f}_{i\rightarrow j},\textbf{f}_{j\rightarrow i}$ on each other, respectively. Due to $\textbf{f}_{i\rightarrow j}+\textbf{f}_{j\rightarrow i}=\textbf{0}$, the net force, resulting from all $i,j$ pairs, vanishes and thence does not contribute to Ohm's law, although the inter-electron coupling will turn out to play a paramount role in the Joule effect.
		\section{The Joule effect}
Because no electron contributes to $j_n$, but the few ones in concentration $2\delta c_n$ with $\epsilon(k)\approx E_F$, showing up as the dashed line in Fig.\ref{o2}, they are also the only ones to be instrumental in the Joule effect. Besides all of them have the same velocity $v_F$. Thus, the well-known formula of the power released by the Joule effect, $\dot {W}_J=\frac{dW_J}{dt}$ ($t$ refers to time), ensues from Ohm's law $j_n=\sigma_n E$, which implies that the friction force equals $2\delta c_n eE$, as
\begin{equation}
\label{joun}
\dot {W}_J=2\delta c_n e\textbf{E}.\textbf{v}_F=\textbf{E}.\textbf{j}_n=\frac{j_n^2}{\sigma_n}\quad,
\end{equation}
for which we have made use of $\textbf{j}_n=2\delta c_n e\textbf{v}_F$.\par
	The Joule effect takes place via two different processes in a superconductor. The calculation of the Joule power ${\dot {W}}_1$, released through process I, is identical to that one leading to Eq.(\ref{joun}) 
\begin{equation}
\label{jou1}
\dot {W}_1=\frac{m}{e\tau_s}\textbf{j}_s.\textbf{v}=\frac{j_s^2}{\sigma_s}\quad,
\end{equation}
where $\textbf{v}$ is the mass center velocity of superconducting electrons ($\Rightarrow \textbf{v}=\frac{\textbf{j}_s}{c_s e}$) and advantage has been taken of Ohm's law in Eq.(\ref{ohm2}) to express the resulting friction force ($=\frac{m}{e\tau_s}\textbf{j}_s$), exerted on the  mass center of superconducting electrons. The physical significance of Eq.(\ref{jou1}) is such that the work $W_1>0$ performed by the driving force $\propto E$ is turned into heat by the friction force.\par
	 However, the calculation of the Joule power ${\dot {W}}_2$, released through process II, proceeds otherwise, because the work $W_2$, to be turned into heat by the friction force, is performed by the \textit{inter-electron} forces, rather than the \textit{driving} one, as seen for $W_1$ in Eq.(\ref{jou1}). Accordingly, while any electron in a normal metal may lose, due to Pauli's principle, an energy randomly distributed from $0$ up to $\epsilon_1-\epsilon_2$ (see Fig.\ref{o2}), conversely the corresponding \textit{internal} energy change, experienced by the BCS electrons, due to the scattering of one electron pair, is \textit{uniquely} defined, as will be shown hereafter.\par	
 In case of $j_s\neq0$, the chemical potential of majority (minority) electrons, characterized by the average velocity $v_s$ ($-v_s$) reads $\mu(c_s+\delta c_s)$ ($\mu(c_s-\delta c_s)$). During each elementary scattering process, a single pair is brought back from the majority subset to the minority one, which results into $\delta \mathcal{E}_s$, the energy lost by the BCS electrons to the lattice, reading
  $$\delta \mathcal{E}_s=\mu(c_s+\delta c_s)-\mu(c_s-\delta c_s)=2\frac{\partial \mu}{\partial c_s}\delta c_s\quad .$$ 
  Since the transfer rate is equal to $2\frac{\delta c_s}{\tau_s}$, the Joule power $\dot{W}_2=2\frac{\delta c_s}{\tau_s}\delta \mathcal{E}_s$ reads finally, due to $\textbf{j}_s=2\delta c_s e\textbf{v}_s$
\begin{equation}
\label{jou2}
\dot{W}_2=4\frac{\partial \mu}{\partial c_s}\frac{\delta c_s^2}{\tau_s}=\frac{j_s^2}{\sigma_J},\quad
\sigma_J=\frac{(ev_s)^2\tau_s}{\frac{\partial \mu}{\partial c_s}}
\quad .
\end{equation}
 The result in Eq.(\ref{jou2}) is noteworthy in two respects :
\begin{itemize}
		\item 
	even though $\dot{W}_2$ is still proportional to $j_s^2$ as $\dot{W}_1$ in Eq.(\ref{jou1}), the conductivity $\sigma_s$, deduced from Ohm's law in Eq.(\ref{ohm2}), \textit{differs} from $\sigma_J$ 
  $$\sigma_s=\frac{c_se^2\tau_s}{m}\neq\frac{(ev_s)^2\tau_s}{\frac{\partial \mu}{\partial c_s}}=\sigma_J\quad ;$$
	\item
	unlike $\sigma_s>0$, the \textit{sign} of $\sigma_J$, which sets whether the Joule heat $W_2$ will flow from the conduction electrons towards the lattice ($\Leftrightarrow W_2>0$), as is always the case in a normal conductor, or conversely will flow into the \textit{reverse} direction ($\Leftrightarrow W_2<0$), is to be determined by the \textit{sign} of $\frac{\partial \mu}{\partial c_s}$. As a matter of fact, the searched criterion for the existence of persistent currents will be worked out by taking advantage of this peculiarity.	
\end{itemize}
\par
It is worth elaborating upon the significance of $\sigma_s\neq\sigma_J$. The Joule power $\dot{W}_J$ reads in general
\begin{equation}
\label{jou3}
\dot{W}_J=\sum_i \textbf{f}_i.\textbf{v}_i\quad ,
\end{equation}
where the sum is carried out on every electron in the conduction band, labeled by the index $i$, moving with velocity $\textbf{v}_i$ and undergoing the friction force $\textbf{f}_i$. Owing to Ohm's law, which implies that the resulting friction force equals $2\delta c_n e\textbf{E}$, and $\textbf{v}_i=\textbf{v}_F$ for all electrons contributing to $\textbf{j}_n=2\delta c_ne\textbf{v}_F$, Eq.(\ref{jou3}) can be recast, for a normal metal, as
$$\dot{W}_J=\sum_i \textbf{f}_i.\textbf{v}_i=\left(\sum_i \textbf{f}_i\right).\textbf{v}_F=2\delta c_ne\textbf{E}.\frac{\textbf{j}_n}{2\delta c_ne}=\frac{j_n^2}{\sigma_n},$$
which is seen to be identical to Eq.(\ref{joun}). Hence the fact, that the \textit{same} conductivity $\sigma_n$ shows up in both expressions of Ohm's law $j_n=\sigma_n E$ and the Joule effect $\dot{W}_J=\frac{j_n^2}{\sigma_n}$, is realized to result from the typical property of a degenerate Fermi gas, that all electrons, contributing to $\textbf{j}_n$, have the same and one velocity $\textbf{v}_F$. Besides, Eq.(\ref{joun}) expresses also the fact that the Joule heat $W_J$ is equal to the work performed by the driving force $\propto E$. However this is no longer true for the BCS state, because of the additional contribution $W_2$, expressed in Eq.(\ref{jou2}). Accordingly, in case of a BCS state, the whole Joule power, reads as  
	$$\dot{W}_J=\dot{W}_1+\dot{W}_2=j_s^2\left(\sigma_s^{-1}+\sigma_J^{-1}\right)\Rightarrow W_J\neq W_1\quad .$$\par
Finally it remains to be shown that $W_2=0$ in a normal metal. The time-derivative of the work done by the inter-electron forces $\textbf{f}_{i\rightarrow j},\textbf{f}_{j\rightarrow i}$ reads $\dot{W}_{ij}=\textbf{f}_{i\rightarrow j}.\textbf{v}_j+\textbf{f}_{j\rightarrow i}.\textbf{v}_i$. Meanwhile $\textbf{f}_{i\rightarrow j}+\textbf{f}_{j\rightarrow i}=\textbf{0}$ and $\textbf{v}_i=\textbf{v}_j=\textbf{v}_F$ imply that $\dot{W}_{ij}=0$. \textit{Q.E.D.}
	\section{Prerequisite for the existence of persistent currents}
	The applied field $E$ gives rise to the total current $j=j_n+j_s$, where $j_n=\sigma_nE$ and $j_s=\sigma_sE$, as required by Ohm's law. After $E$ has vanished, $j_n$ is quickly destroyed by the Joule effect. However whether $j_s$ will decay down to $0$ or conversely will turn to a persistent current, will be shown hereafter to depend \textit{solely} upon the \textit{sign} of the \textit{whole} Joule power $\dot{W}_J$, generated via processes I and II. In case of $E=0$, the kinetic energy, associated with $j_s\neq 0$, $\mathcal{E}_K=\frac{c_sm}{2}v^2=\frac{m}{2c_se^2}j_s^2$, due to $j_s=c_sev$, is turned into heat by the friction force. The expression of $\dot{\mathcal{E}}_K$ is obtained, thanks to $j_s=2\delta nev_s$ and $\dot{v}=-\frac{2\delta nv_s}{c_s\tau_s}$, as
	$$\dot{\mathcal{E}}_K=-c_smv\dot{v}=-\frac{m}{c_se^2\tau_s}j_s^2=-\frac{j_s^2}{\sigma_s}\quad ,$$
	so that the expression of $\dot{W}_1=-\dot{\mathcal{E}}_K$ remains unaltered with respect to that one in Eq.(\ref{jou1})  and finally we get the same expression as in the $E\neq 0$ case, i.e.
		 $$\dot{W}_J=j_s^2\left(\sigma_s^{-1}+\sigma_J^{-1}\right)\quad.$$\par
	  If $\dot{W}_J>0$, the Joule effect will cause eventually $j_s=0$ and the associated kinetic energy will be converted into heat, to be dissipated in the lattice, as occurs in a normal metal. Inversely in case $\dot{W}_J<0$, which requires \textit{both} $\sigma_J<0\Leftrightarrow\frac{\partial \mu}{\partial c_s}<0$ (see Eq.(\ref{jou2})) \textit{and} $\sigma_J+\sigma_s>0$, the Joule heat is seen to be bound to flow from the lattice \textit{towards} the superconducting electrons, which will cause the lattice temperature to \textit{decrease}. However, since such a spontaneous cooling of the system, comprising \textit{all of electron and lattice degrees of freedom}, which can furthermore \textit{exchange neither heat, nor work with the outer world} due to $E=0$, would cause its  \textit{whole entropy} to \textit{decrease}, and would thence be tantamount to \textit{violating the second law of thermodynamics}, the searched criterion is deduced to say that persistent currents can be observed, \textit{only} if both following conditions are fulfilled
\begin{equation}
\label{criter}
\frac{\partial^2 \mathcal{E}_s}{\partial c_s^2}=\frac{\partial \mu}{\partial c_s}<0\Rightarrow\sigma_J<0\quad,\quad
\sigma_J+\sigma_s>0\quad.
\end{equation}\par
\begin{figure}
\includegraphics*[height=6 cm,width=6 cm]{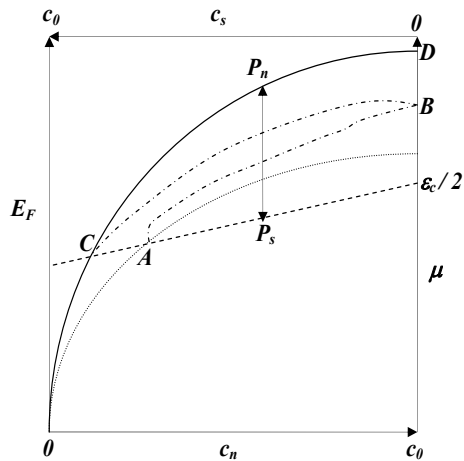}
\caption{schematic plots of $E_F(T_i,c_n)$, $E_F(T_f,c_n)$ and $\mu(c_s)$ as dotted, solid and dashed lines, respectively, in case $\frac{\partial \mu}{\partial c_s}<0$, $\frac{\partial E_F}{\partial c_n}+\frac{\partial \mu}{\partial c_s}>0$; $\frac{\partial \mu}{\partial c_s}$ has been taken to be constant for simplicity; the origin $E_F=\mu=0$ is set at the bottom of the conduction band; the crossing points $A,C$ of $E_F(T_i,c_n),E_F(T_f,c_n)$, respectively, with $\mu(c_s)$, exemplify stable solutions of Eq.(\ref{gidu}); the tiny differences $E_F(T_f,c_n)-E_F(T_i,c_n)$, $E_F(T_i,c_n)-\mu(c_0-c_n)$ have been hugely magnified for the reader's convenience; the dashed-dotted line, linking $A,B,C$ together represents the adiabatic process, discussed in section VI; the points $P_n,P_s$ and the arrow linking them illustrate a superconducting-normal transition \textit{in progress} (i.e. $c_n(T_f)<c_n(P_n)<c_0$, $0<c_s(P_s)<c_s(T_f)$), taking place at $T_f$, under the constraint $c_n+c_s=c_0$}\label{sta}
\end{figure}
\begin{figure}
\includegraphics*[height=6 cm,width=6 cm]{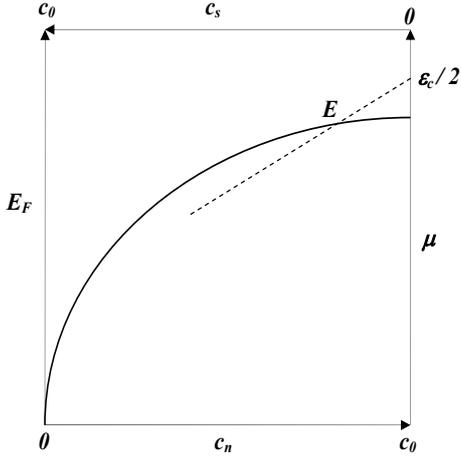}
\caption{schematic plots of $E_F(T,c_n)$ (solid line) and $\mu(c_s)$ (dotted line) in case $\frac{\partial \mu}{\partial c_s}<0$ and $\frac{\partial E_F}{\partial c_n}+\frac{\partial \mu}{\partial c_s}<0$; the crossing point $E$ of $E_F(T,c_n)$ with $\mu(c_s)$ represents an instable solution of Eq.(\ref{gidu})}\label{uns}
\end{figure}

	That those conditions in Eqs.(\ref{criter}) are necessary, but by no means sufficient ones, can be understood by looking back at Eq.(\ref{gidu}). The equilibrium of the mixture of normal and superconducting electrons will be stable provided
	 $$\frac{\partial E_F}{\partial c_n}+\frac{\partial \mu}{\partial c_s}>0\quad .$$
Both stable and instable cases are illustrated in Figs.\ref{sta},\ref{uns}, where $E_F(T,c_n),\mu(c_s)$ have been plotted versus $c_n,c_s$, respectively. Note that $\frac{\partial E_F}{\partial c_n}\approx\rho(E_F)^{-1}>0$ where $\rho(\epsilon)$ is the density of one electron states in the conduction band\cite{ash}. The infinite slope $\frac{\partial E_F}{\partial c_n}\left(c_n\rightarrow 0\right)\rightarrow\infty$ is then typical of a $3$ dimensional van Hove singularity\cite{ash}, associated with the bottom of the conduction band, where $\rho(\epsilon\rightarrow 0)\propto \sqrt{\epsilon}$. The inequality in Fig.\ref{sta}, $E_F(T_i,c_n)<E_F(T_f,c_n),\forall c_n$ with $T_i>T_f$, ensues from $\frac{\partial \rho}{\partial E_F}(E_F)>0$ via the Sommerfeld integral\cite{ash}, which will be shown elsewhere to be another prerequisite for the occurrence of superconductivity. At last in case $c_s\rightarrow0$, there is  $\mathcal{E}_s\approx \frac{\epsilon_c}{2} c_s$ where $\epsilon_c$ refers to the Cooper pair energy\cite{coo}, which entails  that $\mu(0)=\frac{\partial \mathcal{E}_s}{\partial c_s}(0)=\frac{\epsilon_c}{2}$.\par
	The experiment, to be discussed below, is aimed primarily at bringing evidence of the \textit{anomalous} ($\sigma_J<0\Rightarrow\dot{W}_J<0$) Joule effect, associated with a BCS state. Since every superconducting material is claimed here to be characterized by $\dot{W}_J<0$, the experimental procedure will look for evidence of the sample temperature being \textit{lowered} by the Joule effect.\par
	 There are in general two ways to have a current flowing through any conductor, i.e. either directly by feeding an externally controlled, time-dependent current $I(t)$ into the sample, or indirectly by inducing the current $j_s(t)$ via a time-dependent magnetic field $H(t)$ according to Faraday's law\cite{jac}. Though the latter has been overwhelmingly favored\cite{par,tin,way,mei,gen} so far in experiments involving superconductors, the former procedure should be given preference for two reasons :
\begin{itemize}
	\item
as the Meissner effect\cite{sz2} gives rise to a spatially inhomogeneous current $j_s(t,r)$ with $r$ referring to the local coordinate inside the sample, the Joule power $\dot W_J(t,r)$ will thereby vary with $r$, whereas both $j_s(t),\dot W_J(t)$ will remain $r$-independent within the former procedure;  
	\item
because of an irreversible consequence\cite{sz2} of the finite conductivity	$\sigma_s$, there can be no one-to-one correspondence between the applied magnetic field $H(t)$ and $j_s(t,r)$, so that the current distribution remains unknown, by contrast with $j_s(t)=\frac{I(t)}{S},\forall t$ ($S$ refers to the area of the sample cross-section)  within the former procedure.
\end{itemize}\par
	\section{Experimental outlook}
	Consider a thermally isolated, superconducting sample, taken in its initial state $T_i=T(t=0)<T_c, I(t=0)=0,c_n(t=0)=c_n(T_i),c_s(t=0)=c_s(T_i)$ (see $A$ in Fig.\ref{sta}). Then let a direct current $I(t)$ flow through this sample.  $I(t)$ grows from $I(0)=0$ up to its maximum value $I(t_M)$, reached at $t=t_M$, such that $I(t_M)=I_c(T(t_M))$, with $I_c(T(t_M))$ standing for the maximum persistent current\cite{par} at $T=T(t_M)$, which causes the sample to go normal at $t=t_M$, i.e. $c_n(t_M)=c_0\Rightarrow c_s(t_M)=0$ (see $B$ in Fig.\ref{sta}). Then $I(t>t_M)$ decreases from $I(t_M)=I_c(T(t_M))$ back to $I(t_f)=0$, corresponding to the final state, reached at $t=t_f$ and characterized by $T_f=T(t_f),I(t_f)=0,c_n(t_f)=c_n(T_f),c_s(t_f)=c_s(T_f)$ (see $C$ in Fig.\ref{sta}).\par
	The work $W(t_f)$, performed by the external electric field during the thermodynamical process, described hereabove and pictured as a dashed-dotted line in Fig.\ref{sta}, is then given by 
\begin{equation}
\label{wa}
W(t_f)=\int_0^{t_f} U(t) I(t) dt \quad,
\end{equation}
with $U(t)$ designating the measured voltage drop across the sample. Since the sample is thermally isolated, applying the first law of thermodynamics to the system, comprising the independent and superconducting electrons and the lattice, driven from $A$ to $C$ via $B$ through an adiabatic process, yields then
\begin{equation}
\label{one}
Q_2=\int_{T_i}^{T_f}\left(C_\phi(T)+C_s(T)\right)dT-W(t_f)\quad,
\end{equation}
with $W(t_f)$ being defined in  Eq.(\ref{wa}). $C_\phi(T),C_s(T)$ stand for the respective contributions\cite{ash} to the specific heat of the phonons (Debye), which is $I$ independent, and of the conduction electrons, the latter being measured at $T\leq T_c, I=0$. Then the integral over $T$ represents the difference in internal energy of the thermodynamical system, defined above, between $T_i$ and $T_f$. Besides, $Q_2=V\int^{t_f}_{0}\frac{j_s^2(t)}{\sigma_J}dt=\frac{V}{S^2}\int^{t_f}_{0}\frac{I^2(t)}{\sigma_J}dt$ and $V$ are the Joule heat released via process II, and the sample volume, respectively ($j_s(t)$, being $r$-independent, warrants $j_s(t)=\frac{I(t)}{S},\forall t$ and $Q_2\propto V$).\par
 	 As, due to $\sigma_J<0$ and $\sigma_J+\sigma_s>0$ (see Eqs.(\ref{criter})), the Joule effect is expected to \textit{cool down} the sample, we predict that Eq.(\ref{one}) will be fulfilled with $Q_2<0$ and $\frac{dT}{dt}\left(t\in\left[0,t_f\right]\right)<0\Rightarrow T_f<T_M<T_i$, in \textit{full agreement} with a remark by De Gennes\cite{gen} (see\cite{gen} footnote in p.18): \textit{if one passes from the superconducting state to the normal one in a thermally isolated specimen, the temperature of the sample decreases}. Although this experiment could be done as well with $I(t_M)<I_c(T(t_M))$, the condition $I(t_M)=I_c(T(t_M))$ secures the largest $T_i-T_f$, because it maximizes $\left|j_s(t)\right|$ and thence $\left|\dot W_2\right|$.\par
	   Furthermore, the low value of $T_c$, encountered in first kind superconductors, ensures that $C_s(T\leq T_f)$ is known accurately. Conversely, for second kind superconductors, which includes all high-$T_c$ compounds, $C_s(T)$ is negligible\cite{ash} with respect to $C_\phi(T)$, so that Eq.(\ref{one}) gets simpler 
\begin{equation}
\label{two}
Q_2\approx\int_{T_i}^{T_f}C_\phi(T)dT-W(t_f)\quad .
\end{equation}
Due to $C_\phi(T)$ being $I$ independent, unlike $C_s(T)$, taking the time derivative of Eq.(\ref{two}) yields in addition
		$$\frac{V}{S^2}\frac{I^2(t)}{\sigma_J(t)}=C_\phi(T)\frac{dT}{dt}-U(t)I(t)\quad ,$$
which enables one to assess $\sigma_J(t)<0$ for $t\in\left[0,t_f\right]$ and thence to check $\sigma_J(t)+\sigma_s(t)>0$, the necessary conditions for the existence of persistent currents (see Eqs.(\ref{criter})), provided $\sigma_s$ has been measured independently\cite{ges,sar} (the $t$ dependences of $\sigma_s=\frac{c_s e^2\tau_s}{m}$ and $\sigma_J=\frac{(ev_s)^2\tau_s}{\frac{\partial \mu}{\partial c_s}}$ are both mediated by the $j_s$ dependence of $c_s(t)$, as demonstrated hereafter in the concluding section).\par
	Although $T_f<T_i$ entails that the entropy of the two-fluid system decreases, the second law of thermodynamics is thereby not violated, because the electrons remain coupled with the outer world via $I(t)$ during the experiment. At last, note that the state, illustrated by $B$ in Fig.\ref{sta}, refers to a metastable equilibrium, because the stable position at $T_f$ is rather inferred to be at $C$ in Fig.\ref{sta}, as required by Eq.(\ref{gidu}). However, were the electron system to go spontaneously from $B$ to $C$, e.g. along the dashed-dotted line, this process would result\cite{sz2} into $\frac{dj_s}{dt}\neq 0$, due to $j_s\neq 0$ at $B$ versus $j_s=0$ at $C$, while the accompanying Joule effect would give rise to a negative entropy variation $\Delta S_{B\rightarrow C}<0$, at odds with the second law of thermodynamics, as noted hereabove.
	\section{Conclusion}
	The anomalous Joule effect is characterized by $\sigma_s\neq\sigma_J$, i.e. the conductivity $\sigma_s$, deduced from Ohm's law, should differ from $\sigma_J$, the conductivity pertaining to the Joule power released through process II.  It ensues \textit{solely} from the \textit{inter-electron} coupling, which causes the BCS electrons to gain the \textit{internal} energy $\delta\mathcal{E}_s=W_2<0$ through process II at the expense of the lattice, while losing simultaneously the \textit{kinetic} energy $W_1>0$ through process I to the lattice, so that $W_1+W_2<0$ gives rise eventually to the cooling effect, embodied by Eqs.(\ref{one},\ref{two}). Due to $W_2=0$ in a normal metal as shown above, the anomalous Joule effect can be observed solely for a many-body bound state, such as the BCS one. Likewise, the existence of persistent currents is warranted as a consequence of $\sigma_J<0$ and $\sigma_J+\sigma_s>0$ (see Eqs.(\ref{criter})), because the resulting Joule dissipation $\dot{W}_J<0$ would run afoul at the second law of thermodynamics, which lends itself to an experimental check, as discussed above.\par
	  Besides, the property $\sigma_s\neq\sigma_J$ implies that Eq.(\ref{gidu}) can \textit{never} be fulfilled in presence of a persistent current $j_s\neq 0$. Here is a proof : consider the electron system in the equilibrium state, defined by $T=T_f,j_s=0$ and represented by $C$ in Fig.\ref{sta}, for which Eq.(\ref{gidu}) is fulfilled. As $j_s$ grows from $0$ up to its maximum value, the electron system shifts away from $C$ : the Fermi gas, represented by $P_n$ in Fig.\ref{sta}, moves, along the solid line, towards $D$, corresponding to the normal state $c_n=c_0\Rightarrow c_s=0$, while the BCS state, represented by $P_s$, goes, along the dashed line, towards the single Cooper pair state, characterized by $\mu\left(c_s=0\right)=\frac{\epsilon_c}{2}$, provided the sample remains connected to a heat bath at $T_f$. Meanwhile, whenever the thermodynamical state of the two-fluid system is represented by the pair $\left\{P_n,P_s\right\}$ in Fig.\ref{sta}, Eq.(\ref{gidu}) is no longer fulfilled because of $E_F(T_f,c_n(P_n))>\mu(c_s(P_s)=c_0-c_n(P_n))$, which demonstrates the first order nature of the $j_s$-driven superconducting-normal transition\cite{par,tin,sch,gen}, by contrast with the second order transition, observed at $T_c$ with $j_s=0$, for which Eq.(\ref{gidu}) is indeed fulfilled, i.e. $E_F(T_c,c_0)=\mu(0)=\frac{\epsilon_c}{2}$.\par

\end{document}